\documentclass[onecolumn,letterpaper,amsmath,amssymb,floatfix,aps,preprintnumbers,superscriptaddress]{revtex4}

\usepackage{verbatim}
\usepackage{graphicx}
\usepackage{bm} 
\usepackage{dcolumn}  
\usepackage{epsfig}
\usepackage{color}

\begin{document}

\title{Coulomb interactions between disordered charge distributions}

\author{David S. \surname{Dean}}
\affiliation{Universit\'e de  Bordeaux and CNRS, Laboratoire Ondes et Mati\`ere d'Aquitaine (LOMA), 
UMR 5798, F-33400 Talence, France}
\author{Ali \surname{Naji}}
\affiliation{School of Physics, Institute for Research in
Fundamental Sciences (IPM), Tehran 19395-5531, Iran }
\author{Ronald~R. \surname{Horgan}}
\affiliation{Department~of~Applied~Mathematics~and~Theoretical~Physics,
University~of~Cambridge, Centre~for~Mathematical~Sciences, Cambridge~CB3~0WA, United~Kingdom}
\author{Jalal \surname{Sarabadani}}
\affiliation{Department of Physics, Institute for Advanced Studies in Basic Sciences (IASBS), Zanjan 45137-66731, Iran, and
Department of Physics, University of Isfahan, Isfahan 81746, Iran}
\author{Rudolf \surname{Podgornik}}
\affiliation{Department of Theoretical Physics, J. Stefan Institute, SI-1000 Ljubljana, Slovenia} 
\affiliation{Department of Physics, Faculty of Mathematics and Physics, University of Ljubljana, SI-1000 Ljubljana, Slovenia}

\begin{abstract} 
We review some of the recent results in the context of the Coulomb interaction between dielectric surfaces which are randomly charged but remain net-neutral on the average. 
Such surfaces are found to interact in vacuum with attractive long-range normal forces due to the presence of the quenched charge disorder which can compete with the standard 
Casimir-Van der Waals forces. They can also interact with random lateral forces and torques, which exhibit zero mean values but finite and 
long-range root-mean-square values. These effects can  play an important role in Casimir experiments as well as in the interaction between  
solid surfaces as well as biomolecular surfaces that are often covered by disordered charge distributions. 
\end{abstract}

\maketitle

\section{Introduction}\label{sec1.1}
In most studies of electrostatic interactions between charged bodies a number of simplifying assumptions are made. Beyond the purely geometric simplifications, the charge distribution is often taken to be uniform. This assumption is clearly always an idealization as charge distributions in many systems will be inherently complex and/or disordered \cite{bay11,kim08-1,kim08-2,kim08-3,zag08}.   Examples of charge disorder are common in colloidal and soft matter systems \cite{Rudi-Ali1,Rudi-Ali2,Rudi-Ali3}, specific examples include surfactant coated surfaces  \cite{surf-1,surf-2} and random polyelectrolytes and polyampholytes \cite{ranpol}. Metallic and dielectric surfaces with local dielectric constant variation can also exhibit charge disorder as local variations of the  crystallographic axes of an exposed surface lead to a random surface potential, {\em the patch effect} \cite{kim08-1,kim08-2,kim08-3,zag08,spe03}. Finally the chemical preparation of samples is never perfect and charged impurities abound. The presence of charge disorder, even if the system is overall net neutral, can be shown to have strong effects on the interactions between bodies.  Notably, charge disorder can lead to interactions which can mask the Casimir effect and may play an important role in the Casimir effect experiments, possibly making their interpretation rather delicate \cite{kim08-1,kim08-2,kim08-3}. 

The disorder we will consider in this chapter is defined as {\em quenched} as it is fixed once and for good in the preparation of the system, Fig. \ref{fig1.1}, as in the case of charged impurities which are frozen in the boundaries of the materials and cannot move or react to electric fields acting upon them. It is for this reason that quenched disorder is sometimes referred to as {\em frozen}. In small systems, where the objects are not held fixed and can move and/or rotate with respect to one another, the system will tend to lower its electrostatic energy by aligning and positioning its components appropriately. This effect is believed to be a key component in the so-called {\em lock and key mechanism}, that plays a fundamental role in the biological recognition mechanisms \cite{molec-1,molec-2}. 

In a typical Casimir force experiment where one has a two plate or a sphere-plate configuration, the surfaces in question are held fixed (not free to rotate or laterally to displace with respect to each other) and the two quenched charge distributions will be {\em completely uncorrelated}. If the sphere in a sphere-plate geometry is held close to one part of a large plate, the charge disorder will lead to {\em random} normal and lateral {\em forces} on the sphere, as well as {\em random torques}. If we carry out the same experiment between two different apposed parts of the sphere and the same surface, the fact that the sphere will feel a different charge distribution will lead to different normal and lateral forces and torques, in much the same way as if we had changed the sphere and plates by new ones produced by the same productions process, {\em i.e.}, giving the same statistical disorder. Carrying out a sequence of such experiments will lead to an ensemble of measured forces,  which can then be averaged  to obtain the average force. 
In most cases we will see that the average normal force is non-zero. However, if the charge disorder in the plate is invariant by translation in space, that is to say it looks statistically the same everywhere across the plate, the average translational forces and torques will be zero. However for the translational forces and torques there will be a non-zero variance and they will fluctuate. The amplitude of these {\em sample-to-sample} fluctuations will give us an additional information about the nature of the charge disorder and may be useful in unravelling the various components of the force measured in typical Casimir or other force detection setups \cite{kim08-1,kim08-2,kim08-3}. 

\section{Normal electrostatic forces between charge disordered slabs}
\begin{figure}[t]
\centerline{\psfig{file=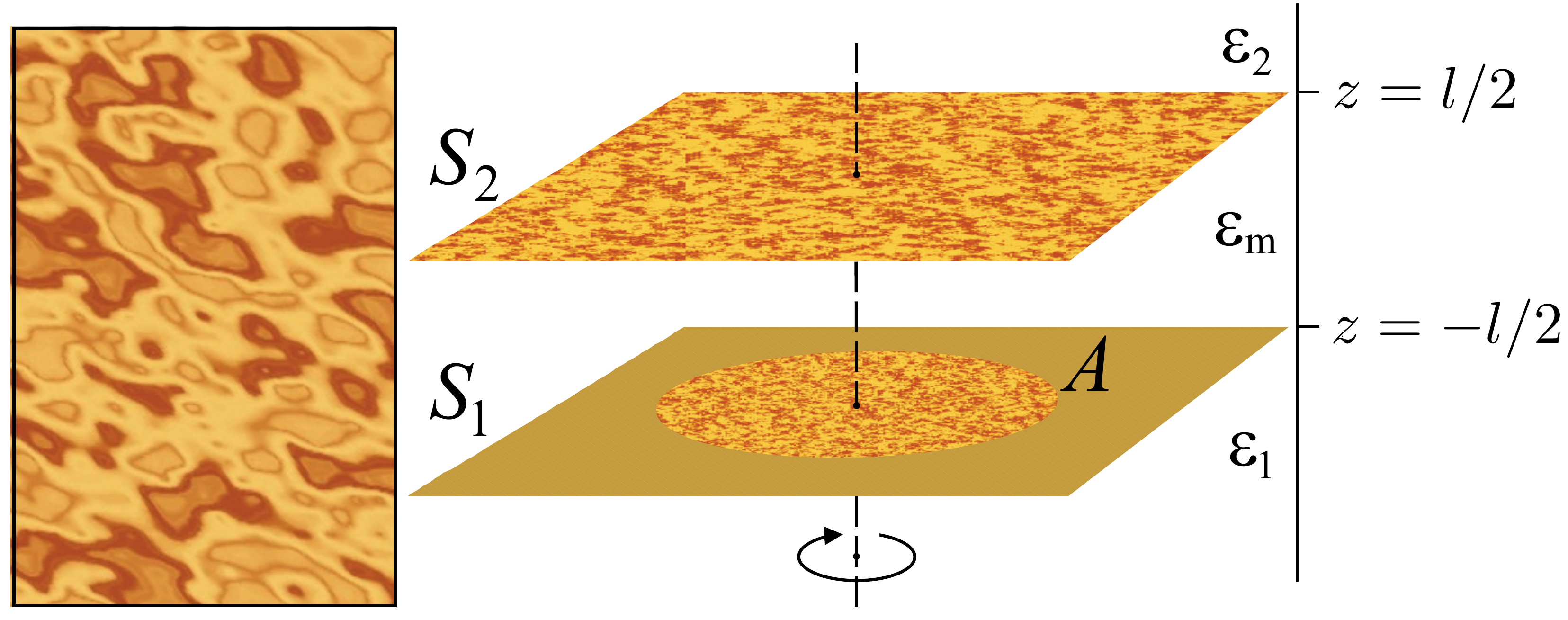,width=8cm}}
\caption{(Left) A schematic quenched surface charge distribution. (Right)  Two parallel dielectric slabs $S_1$ and $S_2$ which are semi-infinite in the $z$ directions and infinite in the ${\bf r}=(x,y)$ plane are placed at a separation distance $l$. The disorder charge  is taken to be distributed over the whole surface in slab $S_2$ but is restricted to the finite region ${\bf r} \in A$ in slab $S_1$.}
\label{fig1.1}
\end{figure}

Consider a system of two parallel semi-infinite dielectrics $S_1$ and $S_2$, with local dielectric constants $\varepsilon({\bf x})$. We take $\varepsilon({\bf x})= \varepsilon_1$ in  $S_1$, $\varepsilon({\bf x})= \varepsilon_2$ in  $S_2$  and $\varepsilon({\bf x})= \varepsilon_m$  in the intervening medium. We denote by $l$ the separation between the two plates and  by  $\rho({\bf x})$ the quenched charge distribution for  a given configuration of the dielectric bodies. The electrostatic energy is given by 
\begin{equation}
E(l) = \frac{1}{2}\int\!\!\!\int d{\bf y} d{\bf x} \,\rho({\bf x}) G({\bf x},{\bf y};l) \rho({\bf y}) 
\end{equation}
where $G({\bf x},{\bf y};l)$ is the Green's function obeying
\begin{equation}
\varepsilon_0\nabla\cdot[\varepsilon({\bf x}) \nabla G({\bf x},{\bf y};l)] =-\delta({\bf x}-{\bf y}),\label{gf1}
\end{equation}
where $\varepsilon_0$ is the permittivity of vacuum. The Green's function depends explicitly 
on $l$ as the overall spatial dielectric function depends on $l$. The average electrostatic energy of this configuration is thus given by
\begin{equation}
\langle E(l)\rangle ={1\over 2} \int d{\bf x} d{\bf y} G({\bf x},{\bf y};l)\langle \rho({\bf x})\rho({\bf y})\rangle,
\end{equation}
where $\langle\cdots\rangle$ indicates the {\em disorder average} over the random charge distributions. The total charge distribution is given by $\rho({\bf x}) = \rho_1({\bf x})+ \rho_2({\bf x})$, where $\rho_{1}({\bf x})$ and  $\rho_{2}({\bf x})$ are  the charge distributions in $S_{1}$ and $S_2$. 
In some cases, notably for the estimation of lateral forces and torques, it is useful to assign one of the bodies, say $S_1$, to have a charge distribution which is restricted to a  subregion of $S_1$ which we will denote by $A$, see Fig. \ref{fig1.1}.

Consider the  case where charge distributions on different bodies are uncorrelated and where the average charge on each body is zero, {\em i.e.} $\langle \rho_1({\bf x})\rangle=\langle \rho_2({\bf x'})\rangle=0$ and  $\langle \rho_1({\bf x})\rho_2({\bf x}')\rangle=0$ for all ${\bf x}\in S_1$ and ${\bf x}'\in S_2$. Then
\begin{equation}
\langle E(l)\rangle ={1\over 2} \int d{\bf x} d{\bf y} G({\bf x},{\bf y};l)[\langle \rho_1({\bf x})\rho_1({\bf y} ) \rangle + \langle \rho_2({\bf x})\rho_2({\bf y}  )\rangle].
\label{enorm}
\end{equation}
We note that the average interaction has just two self interaction terms as there is no interaction on average between charges on different slabs, {\em i.e.}, the interaction energy between a charge in $S_1$ with a randomly chosen charge in $S_2$ is on average zero as the charge in $S_2$ has an equal probability to have either the same sign or the opposite one as that in $S_1$. A non-zero interaction energy is only possible between a charge in a slab and its image charge in the opposing slab, the sign of this image charge depending on the charge in question and the dielectric constants of the system. 

We notice, that there is an interaction between $S_1$ and $S_2$, even if there is no charge in one of the slabs, stemming from an average interaction energy of a randomly charged slab (e.g., $S_1$) with a charge-free slab (e.g., $S_2$). 
The origin of a net interaction between overall neutral surfaces is thus easy to understand, but still rather subtle and surprising. 

We now consider a simple model of the charge disorder, that can originate in either the 
surface charge or bulk charge. On the surface $S_1$ the surface charge distribution is
\begin{equation}
\rho_{1s}({\bf x})= \sum_{i=1}^{N_{1s}} q_i\delta({\bf r}-{\bf r}_i)\delta(z+{l\over 2}),
\end{equation}
where ${\bf r}$ is the two dimensional in plane coordinate.  If the area of the slab  $S_1$ covered with charge disorder is $A$ and the charges have the values $q_i=\pm q_{1s}$ with equal probability and $N_{1s}=n_{1s} A$ where $n_{1s}$ is the surface concentration of charges, we easily find that $\langle \rho_s({\bf r}) \rangle =0$ and
\begin{equation}
\langle \rho_{1s}({\bf x})\rho_{1s}({\bf x}')\rangle = g_{1s}\delta(z+{l\over 2})\delta(z'+{l\over 2})\delta({\bf r}-{\bf r}'),
\end{equation}
where $g_{1s} = n_{1s}q_{1s}^2$. Typical values of $q_{1s}$ are given by the 
electron charge $e$ and the bulk impurity charge densities are between
$10^{10}$ to $10^{15}$ $e/\rm{cm}^3$ \cite{kao,pitaevskii}. Using the size of a molecular layer
we can thus estimate the typical surface charge density of the surface, generated by cutting the bulk. 
Experimentally the heterogeneous structure of the charge disorder on dielectric surfaces can be measured using  Kelvin force microscopy  \cite{bay11}.

Away from the surface, we assume that bulk charge distribution has the form
\begin{equation}
\rho_{1b}({\bf x})= \sum_{i=1}^{N_{1b}} q_i\delta({\bf x}-{\bf x}_i),
\end{equation}
where the charges $q_i$ take the value $\pm q_{1b}$ with equal probability and have concentration
$n_{1b}$. The bulk correlation function is then
 \begin{equation}
\langle \rho_{1b}({\bf x})\rho_{1b}({\bf x}')\rangle = g_{1b}\delta(z-z')\delta({\bf r}-{\bf r}'),
\end{equation}
where $ g_{1b}= n_{1b}q_{1b}^2$. The disorder assumed here has a short range correlation, an assumption easily modified to more general forms \cite{naj10,sar10,last12}. To give explicit formulas we will continue with short range correlation functions and assume that the (infinite) surface areas $S_2=S_1=S$ (by taking $A$ to infinity in the case of slab $S_1$) 
are completely covered by charge disorder. From Eq. (\ref{enorm}) we find that 
\begin{eqnarray}
&\langle E(l)\rangle& = {S\over 2}  [g_{1s}G({\bf 0},-{l\over 2},-{l\over 2};l)  +
 g_{2s}G({\bf 0},{l\over 2},{l\over 2};l)  ]\nonumber \\
 &+& {S\over 2} \int^{-l/2}_{-\infty}\!\!\!\!\!\!\!\!dz\  g_{1b} G({\bf 0},z,z;l)  + {S\over 2} \int^{\infty}_{l/2}\!\!\! dz \  g_{2b}G({\bf 0},z,z;l) ~~~~~~
\label{enorm_2}
\end{eqnarray}
The  Fourier transform in the in-plane coordinates is  
\begin{equation}  
 \tilde G({\bf k}, z, z';l)= \int d{\bf r} \ G({\bf r}, z,z';l) \exp(-i{\bf k}\cdot {\bf r})
 \end{equation}
 and we can use equation (\ref{gf1}) to show that 
 \begin{equation}
 \varepsilon_0{d\over dz}\varepsilon(z){d\over d z} \tilde G({\bf k}, z,z';l) -
 \varepsilon_0\varepsilon(z)k^2 \tilde  G({\bf k}, z,z';l)=
 -\delta(z-z')
 \end{equation}
 where $k = |{\bf k}|$. The above equation may be solved and we find that, for instance, when 
 $z\leq -l/2$ we have 
 \begin{equation}
 \tilde G({\bf k}, z,z;l) = {1\over 2\varepsilon_1\varepsilon_0 k}\left(1 + {[\Delta_1 \exp(kl) -\Delta_2\exp(-kl)]\exp(2kz)\over 1-\Delta_1\Delta_2\exp(-2kl)}\right),
 \end{equation}
where $\Delta_a$ (for $a=1$ or $2$) is the {\em dielectric jump parameter}
\begin{equation}
\Delta_a ={ \varepsilon_a -\varepsilon_m\over  \varepsilon_a +\varepsilon_m}.
\end{equation}
 With this result we can evaluate the terms appearing in equation (\ref{enorm}) using the inverse
 two dimensional  Fourier transform 
 \begin{equation}
 G({\bf 0}, z,z;l) = {1\over (2\pi)^2}\int d{\bf k} ~\tilde G({\bf k},z,z;l).
 \end{equation}
For short range disorder, all the integrals can be evaluated analytically. It turns out that some integrals diverge due to self energies that are independent of the separation of the plates and therefore do not contribute to the force. The total average force can be decomposed into the contribution coming from each of the charge distributions in the system. For instance the average normal force 
$\langle f^{(n)}_{s1}\rangle$ due to the random surface charge on slab $1$ is given by,
 \begin{equation}
{ \langle f^{(n)}_{s1}\rangle \over S}= {g_{1s}(1-\Delta_1^2)\over 16 \pi \Delta_1 \varepsilon_1\varepsilon_0  l^2}\ln(1-\Delta_1\Delta_2)
 \label{surff}
 \end{equation}
 and the contribution of the force from the surface charge on slab 2 is obtained by switching the 
 index $1$ for $2$ an vice a versa in Eq. (\ref{surff}).  
 
 We first note that the interaction is an unscreened $1/l^2$ power law. Furthermore we see that the sign of the interaction is rather nontrivial and depends on the signs  of the dielectric jump parameters $\Delta_a$ \cite{naj10,sar10,dea11}.  We can make the following observations for, e.g., the force due to the slab 1, $f^{(n)}_{s1}$: (i)  for $\Delta_1 >0$ and $\Delta_2>0$,  we find that 
 $\langle f^{(n)}_{s1}\rangle<0$ (attraction), (ii) for  $\Delta_1 >0$ and $\Delta_2<0$,  $\langle f^{(n)}_{s1}\rangle>0$ (repulsion) (iii) for $\Delta_1<0$ and $\Delta_2 >0$,   $\langle f^{(n)}_{s1}\rangle < 0$ (attraction) and (iv) for $\Delta_1<0$ and $\Delta_2 <0$,   $\langle f^{(n)}_{s1}\rangle > 0$ (repulsion). So we see that the surface charge on slab $1$ feels an attractive force toward slab $2$ if $\Delta_2 >0$, {\em i.e.}, if the material composing slab $2$ is more polarizable than the intervening medium. In the case of $\Delta_2>0$ but $\Delta_1<0$, the force due to the charge on slab $1$ is attractive, but due to the charge on slab $2$  is repulsive!
 It is interesting to note that the sign of the thermal van der Waals interaction between two dielectric slabs depends only on the product $\Delta_1\Delta_2$ as we shall see below.
   
The normal force due to the bulk charge disorder in slab $1$ can be shown \cite{naj10} to be given by

 \begin{equation}
 {\langle f^{(n)}_{b1}\rangle \over S}= -{g_{1b} \Delta_2(1-\Delta_1^2)\over 16 \pi  \varepsilon_1\varepsilon_0  l}{1\over 1-\Delta_1\Delta_2}.
 \label{surff2}
 \end{equation}
The result for the force due to the charge distribution  in the slab  $S_2$ is obtained on  switching indices.  The sign of the force depends purely on that of $\Delta_2$ ($\Delta_1$) andbehaves in the same way as the surface charge. However, we see that the force generated by bulk charge disorder is longer ranged than due to the surface charges, decaying as $1/l$. 
In the above computation it is important to bear in mind that slabs will have a finite thickness--changing the inter slab separation should not introduce/remove quenched  charge from the system, which will give an effective bulk term in the energy. A way of avoiding this, without complicating the dielectric problem, is to assume that  the quenched charge disorder has a support which is in the interval $[-(L+l)/2, -l/2]$, in slab 1 for instance, while keeping $L$  fixed upon changing $l$. 

The components of the force due to both bulk and surface charge disorder in the same slab have the same sign, this means that the electrostatic component of the normal force is monotonic, having the same sign for all values of $l$. If one takes into account thermal Casimir forces which are also monotonically attractive or repulsive, it is possible to find systems where the overall force becomes zero, corresponding to a stable or unstable equilibrium \cite{sar10,last12}. The normal thermal Casimir or the zero Matsubara frequency van der Waals (vdW) force interaction behaves as 
\begin{equation}
{f^{(n)}_{\rm{vdW}}\over S} = -{k_B T \, {\rm Li}_3(\Delta_1\Delta_2)\over 8\pi l^3},
\end{equation}  
where ${\rm Li}_3(z)$ is the trilogarithm function. If the prefactor is chosen to be negative (by taking
$\Delta_1\Delta_2 <0$) the repulsive vdW interaction can stabilize the interaction, preventing collapse at short distances. It is theoretically possible to have an attractive charge disorder interaction, behaving as $1/l$ or $1/l^2$ at large separations, meaning that the overall interaction potential will possess a stable minimum  \cite{sar10,last12}.    

The above results can be easily generalized to the case where the charge disorder possesses a 
finite correlation length in the plane of the slabs  \cite{naj10,sar10}. However the case of a non-zero correlation 
normal to the slabs, in the $z$ direction, is more difficult to deal with.  

Finally we should note that the physical usefulness of an average value depends on the scale of
fluctuations of the force with respect to the average force. For simplicity we will consider the 
variance of the normal force with just surface charge disorder. After a rather long computation for
surface charge disorder with zero-correlation length, we find the variance of the {\em total} interaction force induced by surface charge disorder to be given by \cite{dea11}
\begin{equation}
\langle \big({f^{(n)}_s}\big)^2 \rangle_c = {S\over 4\pi\varepsilon_0^2 \varepsilon_m^2 l^2}\left( g_{1s}^2 D_{11} + g_{2s}^2 D_{22} +
2g_{1s}g_{2s}D_{21}\right)
\end{equation}
where 
\begin{equation}
D_{ij} \Delta_i \Delta_j = { \varepsilon_m^4 \over 3  (\varepsilon_m+\varepsilon_i)^2(\varepsilon_m+\varepsilon_j)^2}\left[ {2{\Delta_1\Delta_2}\over  (1-\Delta_1\Delta_2)^2} + \alpha_{ij} {\ln(1-\Delta_1\Delta_2)}\right] 
\label{eq:D_ij}
\end{equation}
where we have define $\alpha_{ij}=(1+\delta_{ij})(-1)^{i+j}$ for $ i, j = 1,2$ (note that no summation  over the indices $i, j$ is assumed in Eq. (\ref{eq:D_ij})). The normal force fluctuations due to surface charge disorder then scale as 
\begin{equation}
\delta f^{(n)}_s \sim {\sqrt{S}\over l}.
\end{equation}
The ratio of the force fluctuations to the average force $ R= \delta f^{(n)}_s/ \langle f^{(n)}_s\rangle$ thus scales as  $R\sim l/\sqrt{S}$ and  are small as the surface area is increased (as expected from self averaging in the thermodynamics limit), however they increase as the inter-slab separation $l$ is increased. In most Casimir type experimental  setups the linear dimensions of the interacting objects are typically much greater than their separation, keeping the force fluctuations small \cite{kim08-2}. 

\section{Lateral electrostatic forces between charge disordered slabs}

Suppose now that the region containing the charge disorder in the dielectric slab 1 is laterally smaller than the slab 2 whilst remaining large of area $A$, see Fig. \ref{fig1.1}. 
We furthermore assume that charge disorder is distributed only on the bounding surfaces of the two slabs. 
If we displace slab 1  laterally by a vector ${\bf a}$ within the plane of the slabs, this will induced a change in  the  charge distribution $\delta\rho({\bf x})$ and give a change in energy
\begin{equation}
\delta E = \int d{\bf x} \,d{\bf y}\,\delta\rho({\bf x}) G({\bf x},{\bf y}; l)\rho({\bf y}).
\label{gbeors}
\end{equation}       
Note that in this set up we assume that there is no change in the dielectric function and so the only displaced charges are the quenched ones in slab 1.  If we move slab 1 by an infinitesimal amount ${\bf a}$ it is clear that the only non-zero contribution to $\delta \rho$ is $\delta \rho_1$. The charge density associated with the  slab 1 can be assumed
to consists of point charges $q_i$ at the points ${\bf x}_i=({\bf r}_i, z_i)$ within the slab $1$ (or on its surface in the case or surface charges). If the change in the system is to displace slab 1 by an infinitesimal vector ${\bf a}$, the new charge distribution associated with slab 1 corresponds to
all the charges at ${\bf x}_i$ being displaced to $({\bf r}_i+{\bf a}, z_i)$ when the whole body is displaced by ${\bf a}$.   This then gives
\begin{equation}
\delta \rho({\bf x}) = -{\bf a}\cdot \nabla_{{\bf r}}\rho_1({\bf x}).
\end{equation}
The change in energy on displacing body $i$ through ${\bf a}$ is given only by the interaction of the charges and image charges in slab 1 
 with those in 2, and therefore defines a lateral electrostatic force ${\bf f}$ that allows
us to write the average of this interaction energy as
\begin{equation}
\langle \delta E ({\bf a})\rangle  = -\langle {\bf f}^{(l)}\cdot {\bf a} \rangle =-{\bf a}\cdot  \int d{\bf x} \,d{\bf y}\,\langle \nabla \rho_1({\bf x}) G({\bf x},{\bf y}; l)\rho_2({\bf y})\rangle,
\end{equation}
The average lateral force is thus {\em zero} as we assume that charge distributions on different bodies are uncorrelated.  

Nevertheless the variance of the lateral force will be non-zero and can be computed using the two point correlation function of the charge disorder. For brevity we state only results for surface charge disorder with  spatially uncorrelated charge disorder \cite{dea11}. We find that the lateral force components in the in-plane directions $i, j=1, 2$ have the correlation function
\begin{equation}
\langle f_i^{(l)} f_j^{(l)}\rangle =-{A\delta_{ij}g_{1s}g_{2s} \varepsilon_m^2\over 4\pi \varepsilon_0^2 l^2 (\varepsilon_m + \varepsilon_1)^2(\varepsilon_m+\varepsilon_2)^2\Delta_1\Delta_2}\ln(1-\Delta_1\Delta_2), 
\label{ncrr}
\end{equation}
showing that the typical lateral force fluctuations scale as \cite{dea11}
\begin{equation}
\delta f^{(l)} \sim {\sqrt{A}\over l },
\label{typicalghfdjr}
\end{equation}
the same scaling as for the fluctuations of the normal force.  We note that in order for the variance to be non-zero we do not require any dielectric discontinuities and find that for $\varepsilon_1=\varepsilon_2 = \varepsilon_m$ 
\begin{equation}
\langle f_i ^{(l)}f_j^{(l)}\rangle ={A\delta_{ij}g_{1s}g_{2s} \over 64\pi  \varepsilon_0^2\varepsilon_m^2 l^2}.\label{eqpp0}
\end{equation}

\section{Electrostatic torques between charge disordered slabs}
On rotating the slab 1 by an angle $\theta$ around its symmetry axis \cite{naj12}, that is to say in the direction perpendicular to  the normal between the bounding surfaces of the two dielectric media, we find (assuming that charge disorder is distributed only on the bounding surfaces) that the new charge distribution of slab 1 is given by
\begin{equation}
\rho'_1({\bf x}) = \sum_{n\in S_1}q_n \delta({\bf r}-\hat R_\theta\, {\bf r}_n)\delta(z-z_n), 
\end{equation}
where $\hat R_\theta$ is the two-dimensional rotation matrix. For an infinitesimal rotation angle $\delta\theta$, one has 
$\hat R_{\delta\theta} = 1- \imath (\delta\theta)\,  \hat \sigma_2$, 
where $ \hat \sigma_2=\left( {\begin{array}{cc} 0 & -\imath  \\ \imath & 0  \end{array} } \right)$ is the Pauli matrix.  The only non-zero contribution to $\delta \rho$ in this case 
(assuming the summation over the in-plane Cartesian components $i, j=1, 2$) is then
\begin{eqnarray}
&&\delta\rho({\bf x}) =  \imath(\delta\theta)\,  (\hat \sigma_2)_{ij} r_j \frac{\partial}{\partial r_i}  \rho_1({\bf x}).  ~~~~~~~
\end{eqnarray}
so that the change of the interaction energy, Eq.  (\ref{gbeors}), on rotating the surface $S_1$, is due to the interaction of charges and image charges in $S_1$ with those in $S_2$. We may thus write
\begin{eqnarray}
\delta E &= &  \imath(\delta\theta)\,  (\hat \sigma_2)_{ij}\int 
d{\bf r'}d{\bf r}\,dz\,dz'\,\big[ r'_j \frac{\partial}{\partial r'_i}  \rho_1({\bf r}',z')\big] \quad \qquad \nonumber\\
  &&\qquad \qquad \qquad \qquad\quad \times G({\bf r}-{\bf r}',z,z'; l)\rho_2({\bf r},z),
\end{eqnarray}
where ${\bf r}'$ and ${\bf r}$ are again the two-dimensional coordinates in the planes of $S_1$ and $S_2$ respectively and $z'$ and $z'$  are the coordinates normal to the planes. We again assume that the charge disorder is restricted to a subregion of area $A$ of $S_1$,  see Fig. \ref{fig1.1}. In an experimental set up, it would be more likely that $S_1$ is finite and the charge distribution is over all $S_1$, however, in the approach followed here we avoid dielectric edge effects and the Green's functions involved then retain their planar symmetry.

Integration over the coordinate ${\bf r}'$ is over the finite area $A$, while that over ${\bf r}$ is unrestricted. The torque $\tau$ acting on the surface $S_1$ is then given by $\delta E = -(\delta \theta)\, \tau$. As the charge distribution on the surfaces $S_1$ and $S_2$ are uncorrelated we find that  $\langle \delta E\rangle = -(\delta \theta)\langle\tau\rangle = 0$. Thus the mean torque is zero.  

From here one can derive \cite{naj12} an intuitively clear physical relation between the lateral force fluctuations and the torque fluctuations of the form
\begin{equation}
	\langle \tau^2\rangle = \frac{1}{A}\langle f_i^{(l)}f_j^{(l)}\rangle I_{ij},
\end{equation}
where the moment of inertia tensor is defined as
$$I_{ij} = \int_{A} d{\bf r}\ (\delta_{ij} r^2 - r_i r_j).$$
The torque fluctuations are thus connected with the lateral force fluctuations through a geometric factor encoded by moment of inertia tensor of the region $A$; thus the torque fluctuations are not the same for a square or a disc of the same surface area. This result is completely general and valid for the assumed plane-parallel arrangement of the two disorder-carrying dielectric surfaces.

As the lateral force fluctuations have the scaling $ {\sqrt{A}/ l }$, (\ref{typicalghfdjr}) , we find that
\begin{equation}
\langle f_i^{(l)}f_j^{(l)}\rangle \sim \frac{A}{l^2}.
\end{equation}
It then follows that typical  torque fluctuations  scale as 
\begin{equation}
	\delta\tau  \sim  \frac{A}{l}, 
\end{equation}
being {\em extensive} in terms of the area $A$. This is as one might expect because torque is determined by the geometry of the area $A$ even in the limit of large area (which is not the case for the random lateral force derived above). The geometry-dependence of the torque fluctuations and the scaling with area $A$ are obtained simply from the moment of inertia tensor.

Again we note that in order for the torque variance to be non-zero we do not require any dielectric discontinuities and find that for $\varepsilon_1=\varepsilon_2=\varepsilon_m$ the torque fluctuations are given by
\begin{equation}
	\langle \tau^2\rangle = \frac{g_{1s}g_{2s} A^2}{128\pi^2 \varepsilon_0^2\varepsilon_m^2 l^2}. 
\end{equation}

\section{Lessons}
Charge disorder has important repercussions for electrostatic interactions between net-neutral dielectric slabs bearing random charges on their bounding surfaces and/or in the bulk. Quenched disorder leads to an additive contribution to the net (normal) interaction force that  scales as $1/l$ (or $1/l^2$) for bulk (or surface) charge disorder, may be attractive or repulsive and depends on the dielectric contrast of the materials. 

Because of the nature of electrostatic interactions between disordered media, there is a sample-to-sample variance in the lateral as well as normal forces and it can be substantial. In the ideal limit where the probe area is very large, the average force is much larger than its fluctuations. However, in some experimental setups the probe size may be quite small and the sample-to-sample force fluctuations could become important. In addition, force fluctuations become important at large separations where the normal force is weak. In the case of lateral force variance, since the average is zero, fluctuations are the only thing observable. Fluctuations in the normal and lateral directions are always comparable; for the special case of a uniform dielectric constant the variance of the force fluctuations in the normal direction is exactly twice  the magnitude of the one in the lateral direction \cite{dea11}.

The disorder-induced torque fluctuations scale differently with the surface area of the interfaces bearing charge disorder, with the typical torque fluctuation being {\em extensive} in the surface area $A$. This opens up a feasible way to measure charge disorder-induced interactions between randomly charged media, in a way which is independent of normal force measurements and with a higher signal-to-noise ratio than lateral force measurements.

\bibliography{chapter_disorder-bib} \bibliographystyle{unsrt}

\end{document}